\documentclass{article}

\usepackage{lettrine}
\usepackage{amsmath}
\usepackage{amssymb}
\usepackage{amsthm}
\usepackage{booktabs}
\usepackage{multicol}
\usepackage{multirow}
\usepackage{graphicx}
\usepackage{printlen}
\usepackage{url}
\newcommand{\comb}{\mathrm{III}} 
\newcommand{\ft}[1][]{\mathcal{F}\{#1\}} 
\newcommand{\ift}[1][]{\mathcal{F}^{-1}\{#1\}} 

\title{Sampling Density Compensation using Fast Fourier Deconvolution}
\author{ Rui Luo, Peng Hu, Haikun Qi
\footnote{
Rui Luo is with the School of Biomedical Engineering \& State Key Laboratory of Advanced Medical Materials and Devices, ShanghaiTech University, Shanghai 201210, China (e-mail: \text{luorui2023@shanghaitech.edu.cn}).\\
Peng Hu is with the School of Biomedical Engineering \& State Key Laboratory of Advanced Medical Materials and Devices, ShanghaiTech University, Shanghai 201210, China (e-mail: \text{hupeng@shanghaitech.edu.cn}).\\
Haikun Qi is with the School of Biomedical Engineering \& State Key Laboratory of Advanced Medical Materials and Devices, ShanghaiTech University, Shanghai 201210, China (e-mail: \text{qihk@shanghaitech.edu.cn}).}
}
\date{}

\begin{document}

\maketitle

\begin{abstract}
Density Compensation Function (DCF) is widely used in non-Cartesian MRI reconstruction, either for direct Non-Uniform Fast Fourier Transform (NUFFT) reconstruction or for iterative undersampled reconstruction. Current state-of-the-art methods involve time-consuming tens of iterations, which is one of the main hurdles for widespread application of the highly efficient non-Cartesian MRI. In this paper, we propose an efficient, non-iterative method to calculate DCF for arbitrary non-Cartesian $k$-space trajectories using Fast Fourier Deconvolution. Simulation experiments demonstrate that the proposed method is able to yield DCF for 3D non-Cartesian reconstruction in less than 20 seconds, achieving orders of magnitude speed improvement compared to the state-of-the-art method while achieving similar or slightly better reconstruction quality.
\end{abstract}

\section{Introduction}
\lettrine{N}{on}-Cartesian acquisition is a highly efficient acquisition technique for magnetic resonance imaging (MRI). However, non-Cartesian sampling poses challenges for the reconstruction step. Since for most non-Cartesian sampling patterns, samples at the low-frequency region are much denser than the high-frequency region, direct reconstruction without proper weighting of the non-Cartesian $k$-space may lead to image blurring. By tradition, the weights to balance the sampling density are termed Density Compensation Function (DCF) \cite{pipeSamplingDensityCompensation1999}. DCF is not only essential for straightforward NUFFT reconstruction, but has also been proven to accelerate the convergence in iterative reconstruction and deep-learning reconstruction by improving the conditioning of the problem \cite{pruessmannAdvancesSensitivityEncoding2001}.

A conventional method for calculating the DCF is based on the Voronoi diagram \cite{voronoiNouvellesApplicationsParametres1908}, which is firstly introduced by Rasche et al \cite{rascheResamplingDataArbitrary1999}. In this prior work, Voronoi diagram is employed by intuition: since the area of each cell appears to be smaller when the samples are denser, and the area of each cell can be used directly as the DCF. Although this method is natural and accurate, constructing a Voronoi diagram is extremely computationally expensive and is also numerically unstable for 3D non-Cartesian trajectories.

In 1999, Pipe and Menon proposed to solve the DCF in an iterative manner \cite{pipeSamplingDensityCompensation1999}. The initial guess of DCF is a unity sequence, and in each iterative step, the DCF is convolved by a kernel function $\Psi$. At convergence, the Point Spread Function (PSF) of the sampling would be close to an impulse function, which is also the ideal PSF. Since the effect of a non-uniform sampling of $k$-space can be quantified by the corresponding PSF,  making the PSF close to the ideal impulse function is a feasible way to solve this problem. This method serves as the foundation of most DCF methods proposed afterwards. 

There are several further developments based on the Pipe and Menon's method \cite{pipeSamplingDensityCompensation1999}. The first one is about the choice of the optimal kernel function $\Psi$ by Johnson and Pipe \cite{johnsonConvolutionKernelDesign2009}. In this work, the authors proposed the optimal choice of $\Psi$ in 2D cases is a convolution of two circular binary windows, while in 3D cases the optimal $\Psi$ is the convolution of two spherical binary windows. This method improves the reconstruction accuracy but is still slow for 3D non-Cartesian trajectories. As reported by Johnson et al., it took 350 seconds per iteration to calculate the DCF for a $256^3$ sampling pattern.

Another method is proposed by Zwart et al. \cite{zwartEfficientSampleDensity2012}, in which the authors proposed to replace the direct convolution with a grid convolution. Grid convolution refers to a convolution mapping non-uniform samples to uniform samples, followed by another convolution mapping uniform samples to non-uniform samples. The advantage of this method comes from avoiding the time-consuming searching process in non-uniform to non-uniform convolution. It has been shown that this method can achieve 1.8 seconds per iteration and converge with 40 iterations for a trajectory designed for a matrix size of $100^3$. However, this method is still iterative and lengthy computation time can be expected for a bigger matrix size.

A recent work proposed an optimization approach for calculating the DCF \cite{dworkOptimizationSpaceDomain2023c}, which achieves more accurate reconstruction but increases the runtime by two orders of magnitude compared to the previous iterative method \cite{pipeSamplingDensityCompensation1999}.

In this work, we propose a fast, non-iterative method to calculate DCF using Fast Fourier Deconvolution (FFD), achieving efficient DCF calculation particularly for 3D non-Cartesian sampling patterns.

\section{Theory and Methods}
\subsection{Problem Formulation}
For $k$-space $S_0(\mathbf{k})$, the sampling process can be denoted by:
\begin{align}
    S_1(\mathbf{k})=\comb(\mathbf{k})\cdot S_0(\mathbf{k})
\end{align}
where $\comb(\mathbf{k})$ is an impulse sequence, defined by $\sum_{i=1}^{N_k} \delta(\mathbf{k}-\mathbf{k}_i)$, and $\mathbf{k}_i$ refers to the $i$-th sample in the sampling pattern $\mathbb{K}=\{\mathbf{k}_i\}_{i=1}^{N_k}$, $N_k$ denotes the total number of $k$-space samples.

After applying the density compensation to the sampled $k$-space, we obtain:
\begin{align}
    S_2(\mathbf{k})&=D(\mathbf{k})\cdot S_1(\mathbf{k}) \\
    &=D(\mathbf{k})\cdot \comb(\mathbf{k})\cdot S_0(\mathbf{k})
\end{align}
where $D(\mathbf{k})$ is the density compensation function (DCF). Without loss of generality, $D(\mathbf{k})$ is assumed to be non-zero everywhere.

Then, the weighted sampling pattern (WSP) can be defined as:
\begin{align}
    E(\mathbf{k})=D(\mathbf{k})\cdot \comb(\mathbf{k})
    \label{eq:Wsp}
\end{align}

The point spread function (PSF) $P(\mathbf{x})$ of $E(\mathbf{k})$ is:
\begin{align}
    P(\mathbf{x})=\ift[E(\mathbf{k})]
\end{align}
where $\mathcal{F}$ denotes the Fourier transform, $\mathbf{x}$ represents the spatial domain.

Our goal is to find a $D(\mathbf{k})$ such that $P(\mathbf{x})\approx\delta(\mathbf{x})$ in $\|\mathbf{x}\|<L$ where $L$ denotes the field of view (FOV). $\|\cdot\|$ denotes the $\ell_2$-norm

\subsection{PSF Decomposition}\label{sec:PsfDec}
Let $\hat{E}(\mathbf{k}),\hat{P}(\mathbf{x})$ denote some initial guess of $E(\mathbf{k}),P(\mathbf{x})$ respectively.
$\hat{P}(\mathbf{x})$ can be decomposed into two parts:
\begin{align}
    \hat{P}(\mathbf{x}) &= \hat{P}_\mathrm{in}(\mathbf{x})+\hat{P}_\mathrm{out}(\mathbf{x})
\end{align}
where
\begin{align}
    \hat{P}_\mathrm{in}(\mathbf{x}) &= \hat{P}(\mathbf{x})\cdot W(\mathbf{x}) \notag \\
    \hat{P}_\mathrm{out}(\mathbf{x}) &= \hat{P}(\mathbf{x})\cdot (1-W(\mathbf{x}))
\end{align}
$W(\mathbf{x})$ is a window function such that $W(\mathbf{x})=0$ for all $\|x\|\geq L$.

We can define $\hat{E}_\mathrm{in}(\mathbf{k})$ and $\hat{E}_\mathrm{out}(\mathbf{k})$ as the Fourier pair of $\hat{P}_\mathrm{in}(\mathbf{x})$ and $\hat{P}_\mathrm{out}(\mathbf{x})$, respectively:
\begin{align}
    \hat{E}_\mathrm{in}(\mathbf{k})&=\ft[\hat{P}_\mathrm{in}(\mathbf{x})] \notag \\
    \hat{E}_\mathrm{out}(\mathbf{k})&=\ft[\hat{P}_\mathrm{out}(\mathbf{x})]
\end{align}

According to the linear property of Fourier transform, we have:
\begin{align}
    \hat{E}(\mathbf{k})&=\hat{E}_\mathrm{in}(\mathbf{k})+\hat{E}_\mathrm{out}(\mathbf{k})
\end{align}

It's important to note that unlike $E(\mathbf{k})$, which is an impulse sequence and is zero almost everywhere, neither $\hat{E}_\mathrm{in}(\mathbf{k})$ and $\hat{E}_\mathrm{out}(\mathbf{k})$ is an impulse sequence, and they are non-zero almost everywhere.

\subsection{Optimal Weighted Sampling Pattern}
Using the method noted in Sec.~\ref{sec:PsfDec}, we have
\begin{align}
    \hat{P}_\mathrm{in}(\mathbf{x})&=\hat{P}(\mathbf{x})\cdot W(\mathbf{x}) \notag \\
    \hat{P}_\mathrm{out}(\mathbf{x})&=\hat{P}(\mathbf{x})\cdot (1-W(\mathbf{x})) \notag \\
    \hat{E}_\mathrm{in}(\mathbf{k})&=\ft[\hat{P}_\mathrm{in}(\mathbf{x})] \notag \\
    \hat{E}_\mathrm{out}(\mathbf{k})&=\ft[\hat{P}_\mathrm{out}(\mathbf{x})]
\end{align}

The proposed WSP is given by:
\begin{align}
    E^\star(\mathbf{k})&=\hat{E}(\mathbf{k})/\hat{E}_\mathrm{in}(\mathbf{k}),
    \label{eq:OptWsp}
\end{align}
which yields:
\begin{align}
    E^\star_\mathrm{in}(\mathbf{k})&=\hat{E}_\mathrm{in}(\mathbf{k})/\hat{E}_\mathrm{in}(\mathbf{k})=1 \notag \\
    E^\star_\mathrm{out}(\mathbf{k})&=\hat{E}_\mathrm{out}(\mathbf{k})/\hat{E}_\mathrm{in}(\mathbf{k})
    \label{eq:OptWspIO}
\end{align}

Then, the corresponding PSF is:
\begin{align}
    P^\star_\mathrm{in}(\mathbf{x})&=\ift[1]=\delta(\mathbf{x}) \notag \\
    P^\star_\mathrm{out}(\mathbf{x})&=\ift[\frac{\hat{E}_\mathrm{out}(\mathbf{k})}{\hat{E}_\mathrm{in}(\mathbf{k})}]
    \label{eq:OptPsfIO}
\end{align}

thus, $P^\star(\mathbf{x})\to\delta(\mathbf{x})$ in $\|\mathbf{x}\|<L$ holds only if:
\begin{align}
    P^\star_\mathrm{out}(\mathbf{x})\to 0,\quad 0<\|\mathbf{x}\|<L
\end{align}

The problem of finding the optimal WSP turns into an optimization problem for $W(\mathbf{x})$:
\begin{align}
    W^\star(\mathbf{x})&=\arg\min_{W(\mathbf{x})}\|P^\star_\mathrm{out}(\mathbf{x})\|/P^\star_0,\quad 0<\|\mathbf{x}\|<L
    \label{OptWx}
\end{align}
where $P^\star_0$ denotes the intensity of the impulse at $P^\star(0)$.

\subsection{Min-Max Parameter Search}\label{sec:MinMax}
For practical implementation, we assume a specific form for $W(\mathbf{x})$: $W(\mathbf{x})=1-\|\bar{\mathbf{x}}\|^p$ where $\bar{\mathbf{x}}=\mathbf{x}/L$ and $p$ is the shape parameter to be optimized. Note that $W(\mathbf{x})$ can also be of other forms. We use the expression of $1-\|\bar{\mathbf{x}}\|^p$ because it is tested to be fast for optimization and yields accurate reconstruction as demonstrated in Sec.~\ref{sec:Exp}.

A simple line search can be performed to find the optimal $p$ in $W(\mathbf{x})$. Specifically, by assuming that the optimal $W(\mathbf{x})$ is dimension-independent and trajectory-independent, we can perform a Monte Carlo test by passing in 1D random sampling patterns to obtain the maximum value of the objective function $\|P^\star_\mathrm{out}(\mathbf{x})\|/P^\star_0$ for each $p$. The optimal $p$ can then be chosen to minimize the maximum objective function. This min-max optimization is formulated as:
\begin{align}
    p^\star=\arg\min_{p}\{\max_{i_\mathrm{test}}\|P^\star_\mathrm{out}(x)\|/P_0^\star\}
\end{align}
where $i_\mathrm{test}$ is the index of a Monte Carlo test, $p^\star$ is the optimal shape parameter. For each test, the 1D sampling pattern follows a Gaussian distribution of $\mathcal{N}(0, (k_\mathrm{max}/3)^2)$,where $k_\mathrm{max}$ denotes the maximum of $|k|$.

In conclusion, following the steps above, we can derive the optimal shape parameter $p^\star$, after which we can obtain $W^\star(\mathbf{x})$, $\hat{P}_\mathrm{in}(\mathbf{x})$, $\hat{E}_\mathrm{in}(\mathbf{k})$ sequentially. The optimal weight sampling pattern $E^\star(\mathbf{k})$ is defined by $\hat{E}(\mathbf{k})/\hat{E}_\mathrm{in}(\mathbf{k})$ in Eq.~\ref{eq:OptWsp}, which is essentially a deconvolution recovering $\hat{P}_\mathrm{in}(\mathbf{x})$ to $\delta(\mathbf{x})$, and in practice is done by Fast Fourier Deconvolution (FFD). The relation between $E^\star(\mathbf{k})$ and the optimal DCF $D^\star(\mathbf{k})$ is defined by $E^\star(\mathbf{k})=D^\star(\mathbf{k})\cdot \comb(\mathbf{k})$ in Eq.~\ref{eq:Wsp}.

\section{Experiments}\label{sec:Exp}
In this section, simulation experiments are performed to evaluate the speed and reconstruction quality of the proposed method, compared to the conventional iterative method.

\subsection{Simulation Settings}
\begin{figure}[t!]
    \centering
    \includegraphics{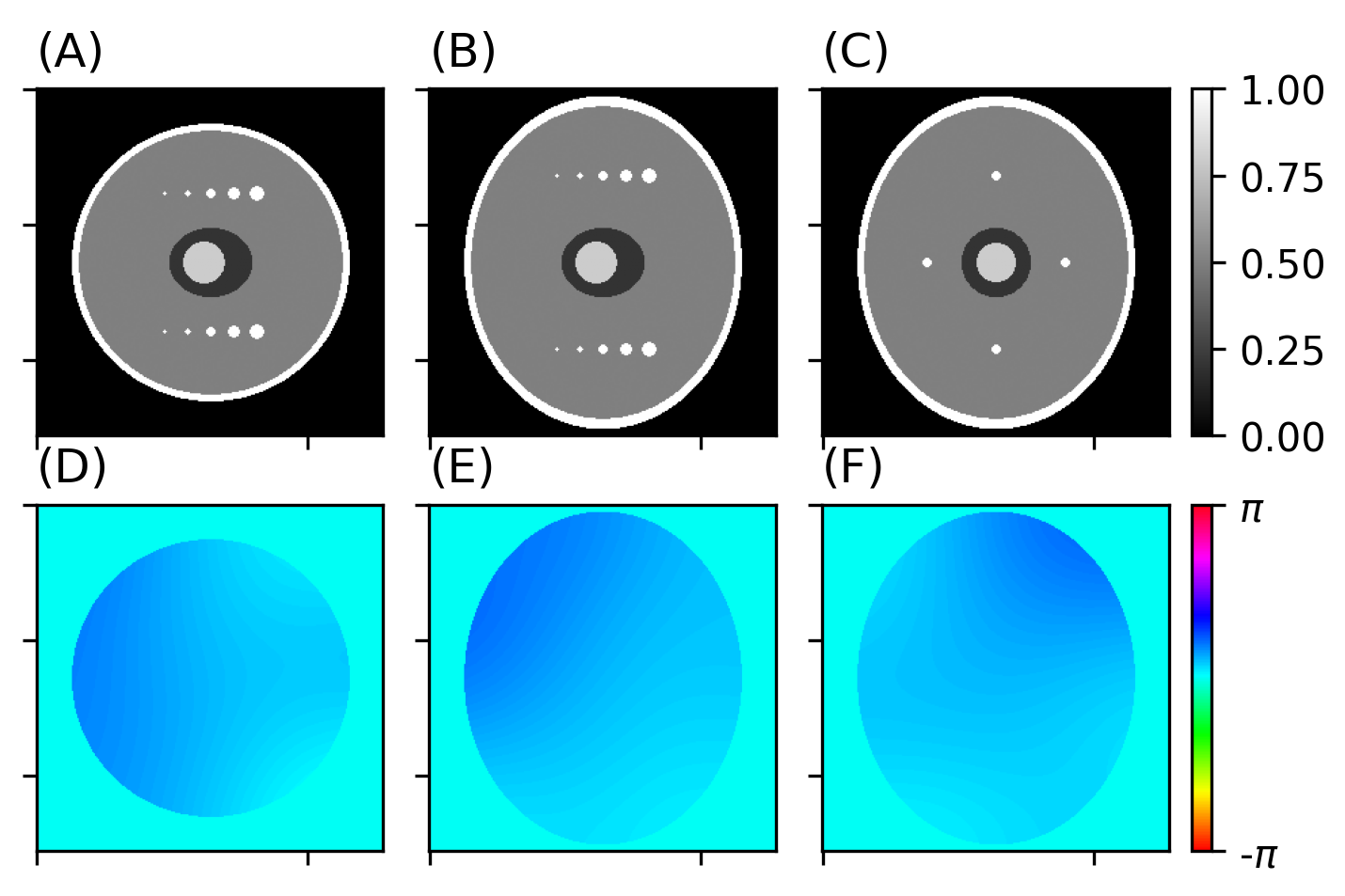}
    \caption{
        The complex-valued digital phantom used for simulation. The top row displays magnitude images of the (A) axial, (B) coronal, and (C) sagittal center slices. The bottom row shows the corresponding phase images for the same slices (D, E, F).
    }
    \label{fig:phantom}
\end{figure}

The simulation is conducted on a complex-valued 2D/3D digital phantom as shown in Fig.~\ref{fig:phantom}, which consists of an elliptical shell, a heart-like structure, and several balls with different sizes. The phase map is generated by sampling the white noise followed by a spatial low-pass filter. The matrix size is set to $256$ and the FOV is set to $500mm$. We test both methods on two 2D trajectories and two 3D trajectories, including Variable Density Spiral \cite{delattreSpiralDemystified2010} (hereafter termed as VdSpiral), Rosette \cite{nollMultishotRosetteTrajectories1997}, Cones \cite{gurneyDesignAnalysisPractical2006} and Yarnball \cite{stobbeThreedimensionalYarnballKspace2021}. FINUFFT \cite{barnettAliasingErrorExpvsqrt1z^22020,barnettParallelNonuniformFast2019} is adopted for non-Cartesian $k$-space simulation and reconstruction. A general-purpose gradient waveform design method \cite{luoRealTimeGradientWaveform2025a} is adopted to simulate the gradient waveforms and the sampling patterns.

\subsection{Implementation Details}
In the proposed method, the 1D DCF is adopted as the initial DCF guess, $\hat{D}(\mathbf{k})$ to maintain numerical stability of the deconvolution, which is defined as:
\begin{align}
    \forall\,i,j\quad \hat{D}(\mathbf{k}_i)=\|\mathbf{k}_{i+1}-\mathbf{k}_i\|_2 \cdot \|\mathbf{k}_i\|_2^{N_d-1},\quad \mathbf{k}_i,\mathbf{k}_{i+1}\in \mathbb{K}_j,
\end{align}
where $\mathbb{K}_j$ is the $j$-th interleaf of $\mathbb{K}$, and $N_d$ is the number of dimension of the $k$-space. $\hat{D}(\mathbf{k})$ of the last point of each interleaf is copied from the previous point. The optimal parameter $p$ of the window function $W(\mathbf{x})$ is obtained by a min-max parameter search in Sec.~\ref{sec:MinMax}.% with $N_\mathrm{test}=100$ and $N_k=256$.

The proposed method was implemented in Python (3.12.8), and FINUFFT was used to perform the deconvolution. The source code of the proposed method will be available upon publication of this work at \url{https://github.com/RyanShanghaitech/MrAutoDcf}.

A state-of-the-art 3D sampling density compensation method by Zwart \cite{zwartEfficientSampleDensity2012}, the C implementation of which is publicly available \footnote{https://www.ismrm.org/mri\_unbound/sequence.htm}, is adopted as the baseline method. This method features the basic iterative structure proposed by Pipe \cite{pipeSamplingDensityCompensation1999}, the optimal kernel proposed by Johnson \cite{johnsonConvolutionKernelDesign2009} and an efficient grid convolution method. It has been demonstrated to achieve a speed of 1.8 seconds per iteration with $100^3$ matrix size \cite{zwartEfficientSampleDensity2012}.

\subsection{Performance Assessment}
Both methods are run on a 4.9~GHz 12-core CPU (Intel® Core™ i7-12700). Normalized Root Mean Square Error (NRMSE) and Structural Similarity Index Measure (SSIM) are calculated for the NUFFT reconstruction using the DCF generated by the baseline and proposed methods to assess reconstruction quality. Execution time $T_\mathrm{exe}$ is recorded to assess the computation speed. The reconstructed images of both methods are normalized to zero mean and unit variance before metric calculation to ensure the comparison reflects structural fidelity and noise amplification, as shown below:
\begin{align}
    \hat{I} = \frac{I - \mu(I)}{\sigma(I)}.
\end{align}

\section{Results}

\begin{figure}[t!]
    \centering
    \includegraphics{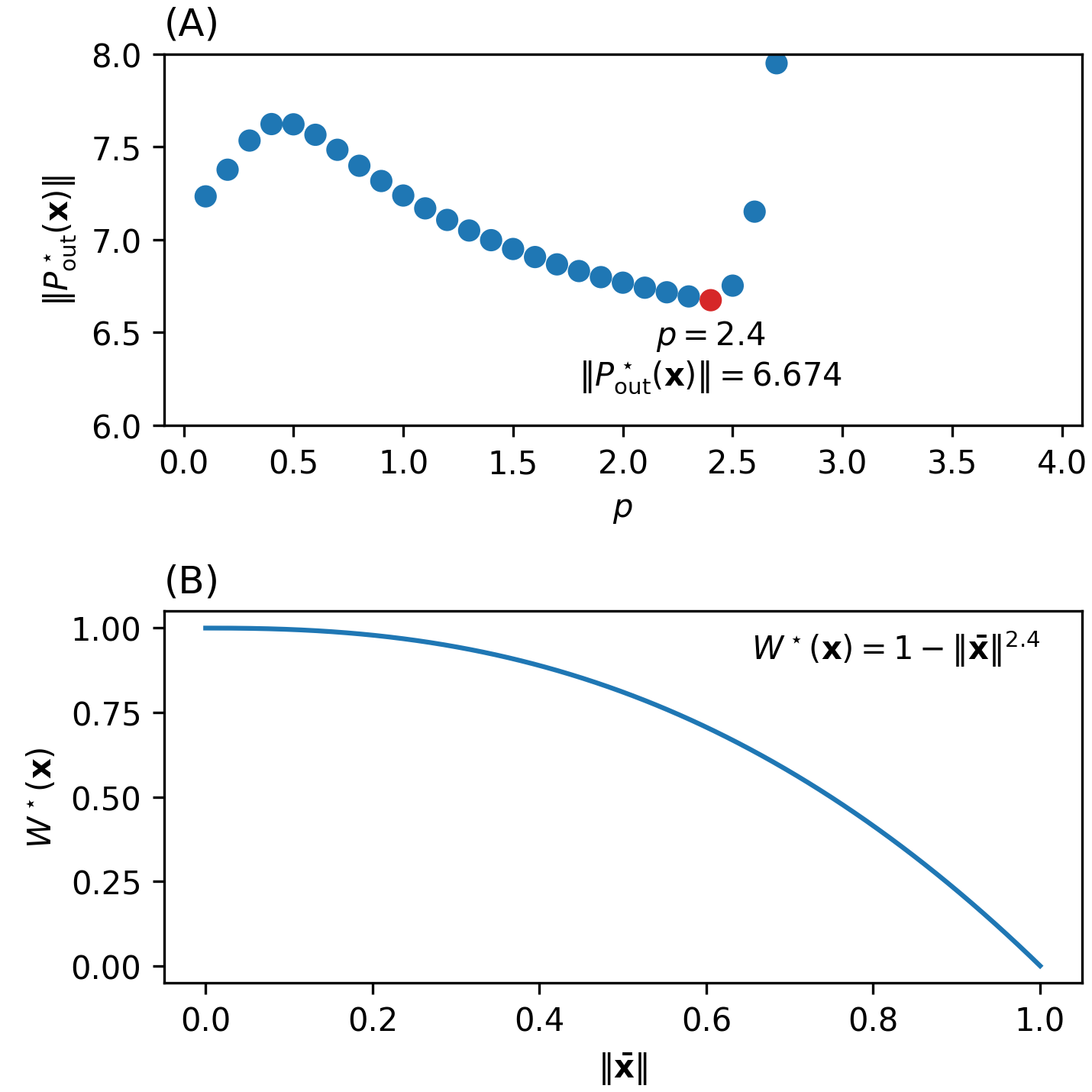}
    \caption{
        (A) Min-max parameter search in the 1D random sampling experiment. (B) The adopted $W^\star(\mathbf{x})$.
    }
    \label{fig:PSearch}
\end{figure}

Among the four involved trajectories, the DCF, the PSF, and the reconstruction results are illustrated for a representative trajectory, Yarnball. Performance metrics including the computation time, NRMSE, SSIM are reported for all four trajectories.

\subsection{Window Function}
The results of the brute-force parameter search of the window function $W^\star(\mathbf{x})=1-\|\bar{\mathbf{x}}\|^p$ are shown in Fig.~\ref{fig:PSearch}. The optimal choice of $p=2.4$ is adopted in the implementation of the proposed method.

\subsection{DCF Results}
\begin{figure}[t!]
    \centering
    \includegraphics{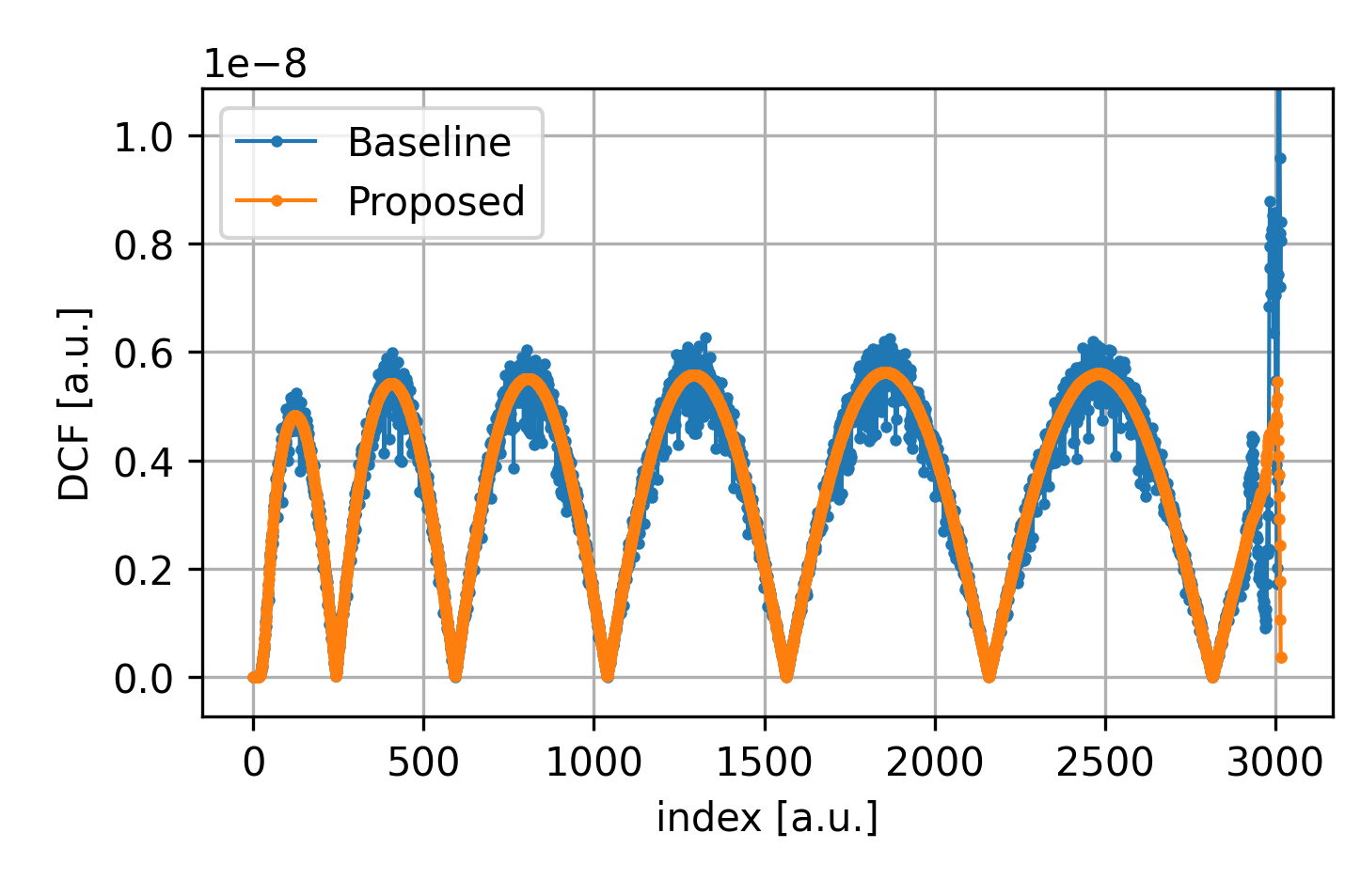}
    \caption{
        DCF calculated using the baseline and the proposed methods for an interleaf of the Yarnball trajectory.
    }
    \label{fig:dcf}
\end{figure}

The DCFs of one representative interleaf of the Yarnball trajectory calculated by the baseline method and the proposed method are shown in Fig.~\ref{fig:dcf}. The two methods yield similar DCFs. However, significant oscillations can be observed in the DCF calculated by the previous method, while the DCF of the proposed method is smooth.  Although the reconstruction results imply that the oscillations do not introduce significant errors in the reconstructed images, a smooth DCF is generally preferred for non-Cartesian MRI reconstruction. It is noted that previous well-established methods also tend to generate smooth DCF, including the Voronoi method \cite{rascheResamplingDataArbitrary1999} and the ad-hoc method such as the analytical DCF of Spiral \cite{hogeDensityCompensationFunctions1997}. 

\subsection{PSF Results}
\begin{figure}[t!]
    \centering
    \includegraphics{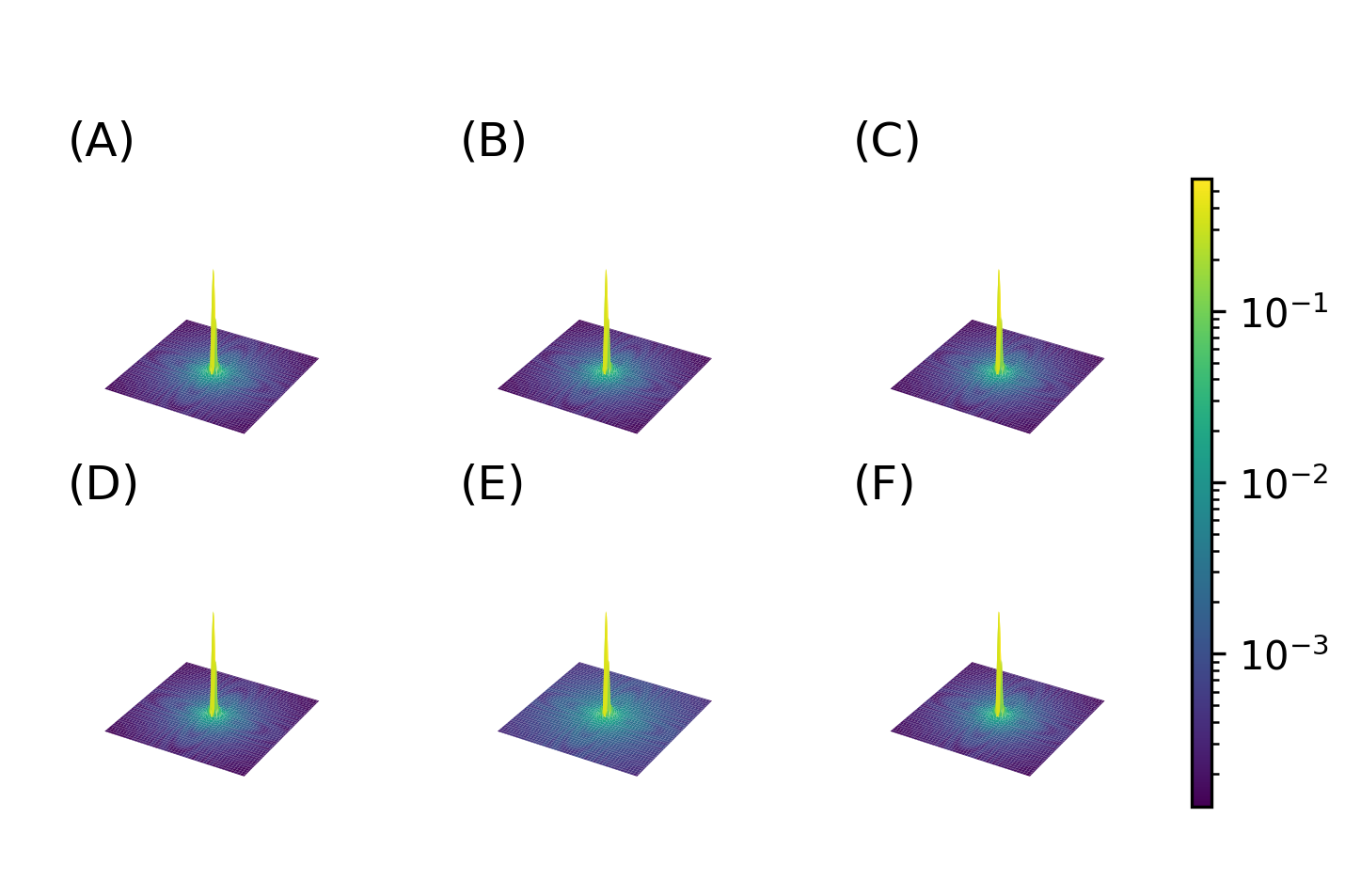}
    \caption{
        PSF resulting from the DCF calculated by the baseline method (A-C) and the proposed method (D-F). The figure displays logarithmic-scaled PSF of three cross-sectional planes:  $Z=0$ (A,D),  $Y=0$ (B,E), and $X=0$ (C,F).
    }
    \label{fig:psf}
\end{figure}

By applying Non-Uniform Inverse Fast Fourier Transform (NUIFFT) to the DCFs calculated by the two methods, we can obtain the corresponding PSFs, as shown in Fig.~\ref{fig:psf}. The full width at half maximum (FWHM) of the PSFs of both methods are $1.5\times$ pixel size, indicating both methods have an equivalent effect on the reconstructed spatial resolution. The PSFs are reconstructed with $10\times$ oversampling for better visualization.

\subsection{Reconstruction Results}
\begin{figure}[p!]
    \centering
    \includegraphics{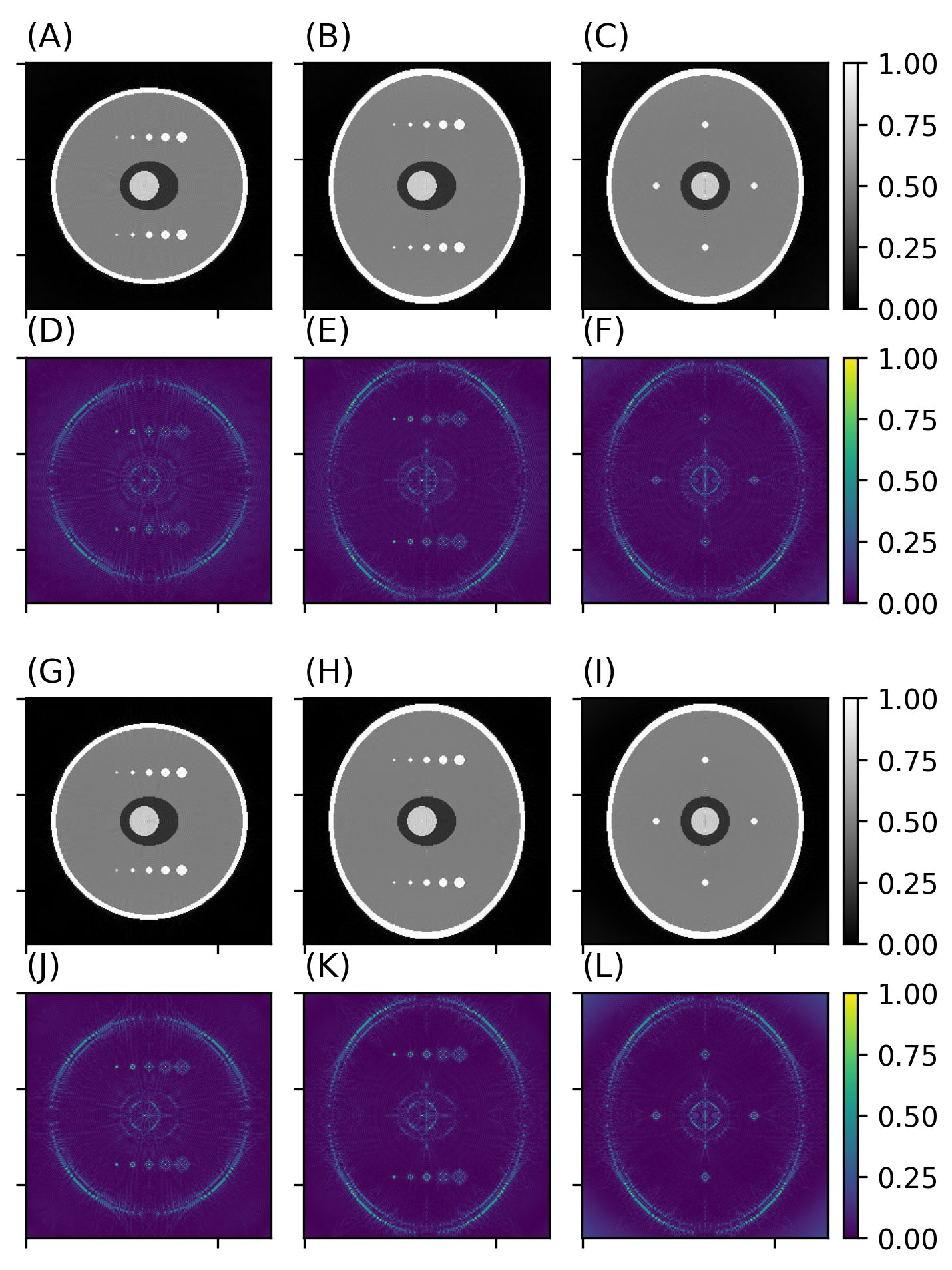}
    \caption{
       The digital phantom images of three orthogonal slices reconstructed using the DCF calculated by the baseline method (A-C) and the proposed method (G-I). The difference images between the reconstructed and ground truth images are shown below the reconstruction for each method. }
    \label{fig:recon}
\end{figure}

The reconstruction results of the 3D digital phantom using the DCF calculated with the baseline method and the proposed method are illustrated in Fig.~\ref{fig:recon}, where the reconstructed images and their difference with the ground truth of three orthogonal slices are shown. Besides the slight Gibbs ringing artifacts due to $k$-space truncation, the reconstructed images are free of noticeable distortion and blurring. The quantitative metrics of the reconstructed images are summarized in Table~\ref{tab:BmQuality}.

\begin{table}[t!]
    \centering
    \begin{tabular}{c|cc|cc}
        \toprule
        ~ & \multicolumn{2}{c|}{NRMSE} & \multicolumn{2}{c}{SSIM} \\
        ~ & Baseline & Proposed & Baseline & Proposed \\
        \midrule
        VdSpiral & 0.018 & 0.016 & 0.953 & 0.956 \\
        Rosette & 0.018 & 0.018 & 0.943 & 0.954 \\
        Yarnball & 0.028 & 0.021 & 0.971 & 0.976 \\
        Cones & 0.023 & 0.019 & 0.971 & 0.976 \\
        \bottomrule
    \end{tabular}
    \caption{Reconstruction quality comparison. The reconstruction metrics of the proposed method and the baseline method are comparable.}
    \label{tab:BmQuality}
\end{table}

\subsection{Speed Comparison}
The execution time of the baseline method and the proposed method is compared in Table~\ref{tab:BmSpeed}. The proposed method is 1-2 orders of magnitude faster than the baseline method while preserving the reconstruction quality as shown in Table~\ref{tab:BmQuality}. Especially, for the 3D non-Cartesian trajectories, the proposed method only takes less than 20 seconds for DCF calculation, facilitating highly efficient 3D non-Cartesian reconstruction.

\begin{table}[t!]
    \centering
    \begin{tabular}{c|cc|c}
        \toprule
        ~ & Baseline & Proposed & Improvement \\
        \midrule
        VdSpiral & 3.835 & 0.044 & 87$\times$ \\
        Rosette & 5.397 & 0.073 & 74$\times$ \\
        Yarnball & 1399.853 & 18.542 & 75$\times$ \\
        Cones & 555.792 & 12.788 & 43$\times$ \\
        \bottomrule
    \end{tabular}
    \caption{Execution time (in seconds) comparison. The proposed method is 1-2 orders of magnitude faster than the baseline method.}
    \label{tab:BmSpeed}
\end{table}

\section{Conclusion}
In this paper, we proposed an efficient DCF calculation method using Fast Fourier Deconvolution. This method is general-purpose, non-iterative and highly efficient. In the simulation experiments, the proposed method reduces the computation time from around 10 minutes to less than 20 seconds for the 3D non-Cartesian trajectories while achieving similar or slightly better reconstruction quality compared to the previous time-consuming iterative method. The proposed DCF calculation method would be a crucial component for an efficient non-Cartesian MRI pipeline.

\section{Acknowledgement}
This work was supported in part by the High Technology Research and Development Center of the Ministry of Science and Technology of China under Grant SQ2022YFC2400133, in part by the Explorer Program of the Science and Technology Commission of Shanghai Municipality under Grant 23TS1400300.

\bibliographystyle{unsrt}
\bibliography{references}

\clearpage

\end{document}